\def\year{2018}\relax
\documentclass[letterpaper]{article} %DO NOT CHANGE THIS
\usepackage{aaai18}  %Required
\usepackage{times}  %Required
\usepackage{helvet}  %Required
\usepackage{courier}  %Required
\usepackage{url}  %Required
\usepackage{graphicx}  %Required
\usepackage{verbatim}

\usepackage{amsmath} % added by Nabil

\begin{document}
\title{Analyzing Uncivil Speech Provocation and Implicit Topics in Online Political News}
\author{Rijul Magu \\Department of Computer Science \\University of Rochester \\Rochester, New York \\rmagu2@ur.rochester.edu
\And
Nabil Hossain \\Department of Computer Science \\University of Rochester \\Rochester, New York \\nhossain@cs.rochester.edu
\And
Henry Kautz \\Department of Computer Science \\University of Rochester \\Rochester, New York \\kautz@cs.rochester.edu
}
\maketitle
\begin{abstract}
Online news has made dissemination of information a faster and more efficient process. Additionally, the shift from a print medium to an online interface has enabled user interactions, creating a space to mutually understand the reader responses generated by the consumption of news articles. Intermittently, the positive environment is transformed into a hate-spewing contest, with the amount and target of incivility varying depending on the specific news website in question. In this paper, we develop methods to study the emergence of incivility within the reader communities in news sites. First, we create a dataset of political news articles and their reader comments from partisan news sites. Then, we train classifiers to predict different aspects of uncivil speech in comments. We apply these classifiers to predict whether a news article is likely to provoke a substantial portion of reader comments containing uncivil language by analyzing only the article's content. Finally, we devise a technique to ``read between the lines'' --- finding the topics of discussions that an article triggers among its readers without frequent, explicit mentions of these topics in its content.
%--- uncovering implicit and latent content within political news articles that trigger new topics of discussions among their reader communities not explicit in the articles themselves.
\end{abstract}

\section{Introduction}
Digital news media has been growing at a rapid pace. The Pew Research Center reported a 20\% increase in the ``Average number of monthly unique visitors for the highest-traffic digital-native news outlets'' between the years of 2014 and 2016~\cite{shearer2017news}. 
% HENRY: Following sentence has to be either cut or replaced with specific facts and citation.
% Also worth considering is the fact that while dependence on  the TV medium for news consumption briskly drops, we observe a spike in consumption of online news.
It is important to study this rich, growing web platform of online news media and to understand the implications of the human connections they forge. For this purpose, we construct a database of articles about the 2016 presidential election to lay the foundations of our analysis. 

\begin{comment}
Not surprisingly, the shift of the user base from newspapers, TV and radio to digital has bolstered attempts of news media companies to cultivate communities of loyal online users.  While theoretically there exists an opportunity of mixed large-scale ideological interactions, the natural conservative-liberal leanings of these media houses often enforce homogeneous. Therefore, we begin our study by estimating whether we can segregate news sources by analyzing the article and individual comment features that identify each source. 
\end{comment}

Furthermore, since the respective audience of these digital platforms aggregate to support a certain set of ideologies, they also contribute significantly to aggressively criticizing the opposing point of view. The toxic environments that form present new opportunities to study the spread of incivility and hatred. Our work focuses on predicting whether an article will generate \emph{substantial} uncivil comments using only the article's content as input.

In general, it is difficult to pinpoint a unanimously agreed upon definition of incivility since it can vary across cultural contexts. Various interpretations of civil and uncivil behavior have been discussed in previous works \cite{papacharissi2004democracy}. For the purposes of this work, we use the definition of incivility provided by \cite{anderson2014nasty} who state incivility to be \textit{``a manner of offensive discussion that impedes the democratic ideal of deliberation''}. 

It is important to note that although the concept of \emph{uncivil speech} overlaps with the concept of \emph{hate speech}, they are not the same thing.  Hate speech ``attacks a person or group on the basis of attributes such as race, religion, ethnic origin, national origin, sexual orientation, disability, or gender'' \cite{hateSpeechDef,davidson2017automated}.  The two concepts coincide, as is often the case, when insulting language is used to attack a protected group or member of a protected group.  Uncivil speech is not hate speech, however, when it is directed toward a non-protected group or its members.  For example, both ``Liberals are perverts'' and ``Trump is an idiot'' are uncivil speech but not hate speech.  On the other hand, hate speech can be expressed in superficially civil language. For example, the statement ``Our nation's social harmony would be improved if persons who did not follow the Christian faith were barred from immigrating'' is hate speech but not uncivil speech. In this paper, any use henceforth of the word ``hate'' means incivility or insult.

An important dimension of digital news media that has escaped mainstream attention is the aspect best described as ``reading between the lines'': do news websites contain latent content that trigger topics of discussion among their readers that are not explicit in their articles' contents? Is it possible for an article to appear neutral in its coverage of an event while invoking biased reactions from its readers? For instance, can an article raise uncivil discussions towards Muslims without mentioning them at all, such as, when simply covering news about terrorism? In our paper, we explore a method to winnow out the latent content in articles through sequential topic modeling analysis of first the articles themselves and then their respective comments.

\section{Related Work}

A number of work in the past have studied the impact of politics on the internet. Adamic and Glance (2005)~\nocite{adamic2005political} analyze the network of liberal and conservative blogs online, and they find an extreme tendency of blogs to link within their respective communities (liberal to liberal, or conservative to conservative). Others have explored the extent of media bias and how it affects the world~\cite{gentzkow2006media,groseclose2005measure}. Researchers have also directly analyzed the relationship between media bias and voting~\cite{dellavigna2007fox}.

Recently, the problem of hate speech has started gaining widespread attention. This is largely due to deeper internet penetration in the past few years and an exponential rise in the use of social media, facilitating the spread of propaganda. Many approaches have been described, most of which use machine learning to predict hateful content~\cite{burnap2015cyber,davidson2017automated}. Others have addressed instances of hate speech where the hate is only indirectly implied in order to get past banned word filters using code-words to represent particular racial slurs~\cite{magu2017detecting}.
%, for example, saying ``Gas the Skypes!'' in place of ``Gas the Jews!'' where ``Skype'' is a code-word for ``Jews''~\cite{magu2017detecting}.

One of the most relevant work for our context is a collaborative work between Google's Jigsaw and the Wikimedia Foundation~\cite{wulczyn2017ex} which classified content on talk pages of Wikipedia along the dimensions of aggression, personal attack and toxicity. We make use of the same Wikipedia dataset, used in their research, to train a classifier that can predict whether an article from our news sources data provokes uncivil language from readers.

\section{Datasets}
\subsection{News Data}
We collected a dataset of articles written on the conservative news website Breitbart and the liberal leaning website Politico for a time period ranging between April 2015 to September 2016, coinciding with the developments towards the 2016 US Presidential Elections. In order to work within the overarching theme of politics, we restricted ourselves to articles containing at least one of the following relevant keywords: \textit{Trump, Clinton, Obama, Hillary, President, Election} and \textit{Immigration}.

We sampled approximately 15,000 articles from both datasets for our analyses. The data contained the articles themselves, along with fields such as date, title, and reader comments.  

\subsection{Wikipedia Detox Project Dataset}
In order to train a classifier that can detect uncivil language, we make use of the publicly available talk page comment corpus made publicly available as part of the Wikipedia Detox project that contains over 100,000 comments by Wikipedia users on talk pages \cite{wulczyn2017ex}. Each comment has been annotated for the following three conditions, each of which can model certain aspects of uncivil conduct:

\begin{enumerate}

\item Toxicity: Does the comment have a healthy or a toxic contribution to the discussion? Does it make you want to leave the discussion or to continue a healthy discussion?
(Annotators rated on an integer scale of 1 to 5, with 3 being neutral)

\item Aggression: Is the comment friendly or aggressive? (Annotators rated on an integer scale of 1 to 5, with 3 being neutral)

\item Personal attack: Does the comment contain a personal attack or harassment? (Annotators were allowed to choose multiple check-boxes, each differing from the others on
the target and the recipient of the attack)

\end{enumerate}

\section{Incivility in News Comments}
To proceed with our analysis of uncivil language, we require the comments in our news dataset to have labels indicating how uncivil they are. The Wikipedia comments dataset provides us with the relevant data for this task since the labels \emph{toxicity}, \emph{aggression}, and \emph{personal attack} are collectively exhaustive in indicating incivility. We create three Logistic Regression classifiers, and using data from this dataset, we train each classifier to identify one of the three aspects of incivility labeled in the dataset. All predictors obtain strong classification performance, as shown by the area under the ROC curve \cite{bradley1997use} in Table~\ref{table:1}.

\begin{table}
\centering
\hfill \break
\begin{tabular}{|c|c|c|c|} 
\hline
     & Toxicity & Aggression & Personal Attack\\ %[0.5ex] 
\hline\hline
    AUC & 0.963 & 0.951 &  0.957\\
    Precision &  0.919 & 0.898 & 0.915 \\
    Recall & 0.590 & 0.548 & 0.552\\
    F1 score &  0.719 & 0.681 & 0.688 \\ \hline
    Accuracy &  95.6\% & 93.4\% & 94.1\%\\
\hline
\end{tabular}
\caption{Performance of Wikipedia Corpus classifiers. AUC is the Area Under the ROC Curve.}
\label{table:1}
\end{table}

\iffalse
\begin{table}
\centering
\hfill \break
\begin{tabular}{|c|c|} 
\hline
     {\bf Classifier} & {\bf ROC AUC}\\ [0.5ex] 
\hline\hline
    Toxicity & 0.963\\
    Aggression & 0.951\\
    Personal Attack & 0.957\\
\hline
\end{tabular}
\caption{Performance of Wikipedia Corpus classifiers.}
\label{table:1}
\end{table}
\fi

We consider a piece of text to consist of uncivil language if it satisfies \emph{any} of the above three conditions. Therefore, for any comment $C$, we define its {\bf incivility score} $I_C$ as the maximum of the comment's aggression, personal attack, and toxicity scores.

%$$I_c = \mfax(Ag,At,T),$$
% where $Ag$, $At$, and $T$, respectively, are the comment's 
% %as predicted by our logistic regression classifiers.
% Note that our incivility score values are real numbers ranging between 0 and 1 since each of these scores are real numbers in the same range. In other words, we do not combine results from the three classifiers to further classify a comments as ``more uncivil'' \emph{vs.}\ ``less uncivil''.

It is important to recognize a limitation of our approach. We expect to obtain a non-trivial number of inaccurate incivility scores for some comments in our dataset. This is because the three classifiers we use were originally trained on a different domain of comments (that of Wikipedia interactions) while our data consists of comments in political news articles. Nevertheless, this is a fast and freely available good alternative to the expensive task of manually annotating the dataset of news comments for uncivil language which we shall explore in future work.

%Note that we simplify the task of finding hate speech in a news comment by , we categorize any comment classified as toxic, aggressive, or personal attack as `hateful'. 

%Therefore, even if a given comment falls within just one of these categories, we would consider that comment to display hate. 

%Hence, it becomes convenient to directly construct the hate label as the  \textit{max} of the three values aggression(Ag), personal attack(At) and toxicity(T) as given in equation~\ref{eqn:1}.

%\begin{equation}\label{eqn:1}
  %Hate=max(Ag, At, T)
%\end{equation}

%\begin{comment}

\section{Articles Provoking Uncivil Speech}

By their very nature, news media websites invite fired-up, exaggerated discussions. This is particularly true for the political context where there is a clear divide within people.
There are  two main reasons why this might happen. First, the ability to have an anonymous online presence removes any social barriers that may otherwise prevent someone from fully expressing their opinion~\cite{rieder2010no}. This provides a fertile ground for trolling. Second, there remains the risk of the emergence of a vicious cycle of sorts beginning by websites attracting audiences that mostly agree with the underlying ideology of the media house. This can further lead to the formation of a vocal community of users who would frequently use the platform. To satisfy the user base, these websites would therefore gravitate further towards the extreme of the political spectrum that this vocal community represents. As this cycle develops, the comments on different political articles by these users can become more and more aggressive. Both these factors contribute in allowing for the spread of incivility on news websites.  

%{\bf Mention classification objective clearly: Therefore, we attempt to predict whether a news article will generate a majority of hateful comments.}

\begin{figure}[t]
\centering
\includegraphics[width=\columnwidth]{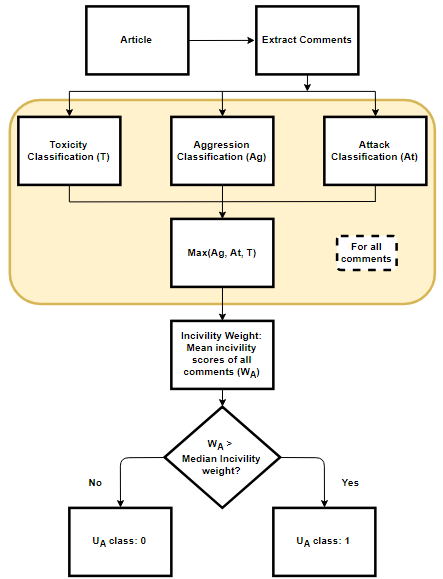}
\caption{Flow chart for labeling ``uncivil speech provoking'' articles.}
\centering
\label{figure:1}
\end{figure}

Our goal is to predict whether a news article will attract an unusual amount of reader comments that contain uncivil language. %For our analysis, we make use of our Hatecom dataset. %Essentially, the objective of the experiment is to evaluate the feasibility of predicting whether a given article (from a particular website) would attract hateful comments.
%{\bf Comments only provide labels, but not used in }
We now describe our sequence of steps to achieve this objective visualized in Figure~\ref{figure:1}.

We randomly sample a set of news articles from Breitbart and from Politico. Then, for any article $A$ having $c_1,c_2,\dots, c_n$ comments, we compute its {\bf incivility weight} $W_A$ as follows:
$$ W_A = \frac{1}{n}\sum_{i=1}^n I_{c_i}$$
Then, for source $S$, we find the {\bf median incivility weight} $M_S$ of all articles sampled from the source.
%Now, given all the articles from a source $S$ (\emph{e.g.,} Breitbart) $a_1,a_2,\dots,a_m$, we calculate the {\bf hate median} $HM_S$ of the source as follows:
%$$HM_S = 
We use this median as a threshold to compute a binary label $U_A$ which is true for an article if the incivility weight of the article is greater than the median.  That is,
\begin{equation}
    U_A=
    \begin{cases}
      1, & \text{if}\ W_A > M_S \\
      0, & \text{otherwise} \nonumber
    \end{cases}
  \end{equation}
In other words, articles with a incivility weight above the source's median incivility weight are labeled ``uncivil speech provoking'' articles. We expect any article predicted to be in this class to trigger a substantial amount of uncivil behavior by commenters of the article. 

$M_{Politico}$ and $M_{Breitbart}$ were found to be 0.089 and 0.081 respectively. It is important to note that the median scores do not necessarily indicate the occurrence of more hate speech in one community in comparison to another. It only informs us about uncivil speech as defined by the classifiers trained on the Wikipedia dataset.

Using median incivility weight as the threshold has the advantage that it avoids class imbalance, since both classes now have equal numbers. In addition, the median, as a relative measure, serves as a replacement of an absolute threshold value in absence of ground truth.  %Additionally, we only require a relative measure within a given context in order to frame a classification problem. 

We next train separate logistic regression classifiers to predict the uncivil speech provoking class of news articles from our two sources. We split our sampled dataset of articles into 80\% for training and 20\% for testing, and we use TF-IDF of bigrams from the article's content data as features. 

\begin{table}
\centering
\hfill \break
\begin{tabular}{|c|c|c|} 
\hline
     &Breitbart&Politico\\ [0.5ex] 
\hline\hline
    ROC AUC & 0.731& 0.690\\
    Precision & 0.622 & 0.674\\
    Recall & 0.661& 0.579\\
    F1 score & 0.641& 0.623\\ \hline
    Accuracy & 65\%& 63\%\\
\hline
\end{tabular}
\caption{Uncivil speech provoking article prediction results.  A random classifier would achieve an accuracy of exactly 50\% because the task is to predict if an article generates more than the median weight of uncivil comments.}

\label{table:3}
\end{table}

\iffalse
\begin{table}
\centering
\hfill \break
\begin{tabular}{|c|c|c|} 
\hline
     &Breitbart&Politico\\ [0.5ex] 
\hline\hline
    ROC AUC & 0.731& 0.690\\
%    Accuracy & 65\%& 63\%\\
    Precision & 0.622 & 0.674\\
    Recall & 0.661& 0.579\\
    F1 score & 0.641& 0.623\\
\hline
\end{tabular}
\caption{Uncivil Speech provoking article prediction results.}
\label{table:3}
\end{table}
\fi
%{\bf mention in defense to table 3 weakness that wikipedia data classification also resulted in low recall --- they do it on comment level, our recalls in table 3 are on an article level}

Results from running the two classifiers are shown in Table~\ref{table:3} which suggest that these classifiers predict whether an article would trigger substantial incivility in its comments much better than chance (50\% accuracy). Further, the results also indicate that based on their linguistic contents, such as the topics they explore and the manner in which they are structured, articles can more or less likely provoke uncivil language from commenters. This has far reaching consequences because it opens the space for studies on the need for news agencies to explore more responsible methods of reporting, so as to minimize the spread of hatred. %From a technical perspective, we expect the performance of these classifier to improve much further if it is trained directly on a human annotated dataset of comments on news articles for hate speech.

\section{Subtext Mining: Finding Implicit Topics}

\begin{comment}

While news sources commonly place themselves on either side of the political spectrum, they are obligated to report professionally. The overarching discourse often involves themes and terminology commonly used among the followers of a specific ideology which might be only sparsely expressed by authors of articles.  
\end{comment}

In this section, our goal is to devise a way to capture latent topics that are relatively restricted in use within article contents but more freely discussed within the community, that is, we wish to extract the \textit{subtext} from the data. 
%We would like to figure out what are the implicit topics in an article that are exposed via reader comments. 
In other words, we would like to figure out: what topics of discussions does an article elicit from its readers that are not explicit during a reading of the article? It is important to note that any analysis directly on the content of the articles would not yield  desired results. In stark contrast, commenters have no incentive to self-censure. In fact, as long as they do not violate abuse policies, site visitors can and do express opinions much more aggressively. Such aggressive remarks not only take the form of strong language, but also use terms and ideas that would be otherwise less frequently appear within the contents of articles. However, these phrases can be mixed with much of the terminology re-used from the articles themselves. Therefore, in order to mine the subtext from the surrounding context, we require a way to segregate content unique to the user base. 

For our analysis, we use the Breitbart dataset of articles. To avoid mixing different themes, we need to filter articles belonging to a common theme. Conveniently, all articles in the data are associated with tags. We focus on news regarding the topic of immigration, and therefore, we select every article which has \textit{Immigration} as one of its tags. 

To segregate the subtext, we carry out topic modeling in the form of Latent Dirichlet Allocation (LDA)~\cite{blei2003latent}. First, we run LDA on the contents of articles in our Breitbart dataset using trigram word features. We collect the top five topics and the associated top five terms (n-grams) for each topic. From this, we extract all unique phrases and store them in a list which we call the {\bf Content Topic Phrases}. Next, we run LDA (using the same conditions) on all the comments associated with articles, passing this list as a list of stop words to ignore. Effectively, we achieve topic phrases unique to commenters that are not explicit parts of the article's contents, thus forming the subtext. We call this set the {\bf Comment Topic Phrases}. The set of these topic phrases are shown in Table~\ref{table:4}.

\begin{table}
\centering
\hfill \break
\begin{tabular}{|p{3.525cm}|p{3.525cm}|} 
\hline
     Content Topic Phrases & Comments Topic Phrases\\ [0.5ex] 
\hline\hline
    white house, illegal aliens, illegal immigrants, obama administration, asylum seekers, law enforcement, federal government, border patrol, united states, immigration reform, resettlement program, paul ryan, executive amnesty, syrian refugees, refugee resettlement program, obama executive, european union, undocumented immigrants, press conference, refugee resettlement & build wall, open borders, american citizens, hillary clinton, catholic church, middle class, civil war, republican party, 3rd world, right wing, good luck,
ted cruz, illegal immigration, white people, cheap labor,
la raza, left wing, political correctness, rule law, western civilization, american people, middle east, saudi arabia, supreme court, democrat party\\
\hline
\end{tabular}
\caption{Topic phrases from subtext analysis on articles related to the topic of immigration.}
\label{table:4}
\end{table}

Immediately, we observe a noticeable difference in the kinds of phrases that emerge from article contents when compared to the comments. The content phrases are straightforward and non-controversial, for example, ``undocumented immigrants'', ``immigration reform'', ``border patrol'', \emph{etc}. However, in the comments we find instances like ``build wall'' (the popular conservative phrase ``build the wall'' implies requests for the construction of a physical wall between the United States of America and Mexico to control immigration) along with ``white people'' and ``western civilization'', probably made in the context of the white race and the western world being under threat from immigrants.
These are instances of ideas often talked about within right wing circles, yet not commonly explicitly discussed in conservative news pieces. Hence, we were able to extract a notion of the subtext of our data. 

\begin{comment}
A frequent occurrence across news sources is the use of topics that are \textit{explicit}, in the sense that they are clearly and openly talked about through the titles and content of the articles themselves. However, there are also topics that are  \textit{implicit}. These topics or terminologies are not explicitly mentioned but are implied within the context. In this paper, we describe an algorithm to be able to extract these latent topics, in other words the subtext of the content.

We carry out topic modeling to achieve this. First, We run LDA on the content of the articles to estimate the top 5 topics and associated top 5 topic phrases. For consistency, we only select a class of articles related to the same topic, which is done by selecting articles under a particular tag. For our analysis, we select all articles marked \textit{Immigration}. Next, we run LDA for the same conditions on the comments of the dataset. The difference between the generated comments reveal the subtext. 

For example, we observe the occurrence of terms such as \textit{speak English} and \textit{build wall} in the results for the comments analysis, but absent in the content analysis. 
\end{comment}

\section{Conclusions}
In this paper, we analyzed online political news using data from conservative (Breitbart) and relatively liberal (Politico) news websites to discover the use of uncivil language within their reader communities. We created classifiers to predict how uncivil a user's comment is on a given news article using three measures of incivility. With the help of incivility labels for comments, we predicted whether a given article would generate substantial incivility within its comment section by only looking at the article's content. The accuracy of this prediction was about 64\%, which is far superior to chance (50\% accuracy) and a satisfactory result given the difficult nature of the task. %and we discussed its potential impact.
Furthermore, we devised a method to mine the subtext from a set of articles, and we applied this method on articles covering the topic of immigration. This allowed us to find implicit, latent content in immigration-related articles via the reader comments. 
There are limitations of our work that can be addressed in future research. We annotated the incivility of our news comments using classifiers trained on data belonging to a non-similar context (Wikipedia comments). This can affect results since comments considered civil in one context can be uncivil in another (and vice-versa). Therefore, creating a thoroughly labeled dataset of news comments using crowd-sourcing and then training a classifier on this dataset will most likely result in a much better performance. 

Applications of our subtext mining model include comparing the degree of implicit topics in articles across different partisan news sources and studying the differences between various ideological communities.

Finally, our approaches for detecting uncivil speech provoking content and for subtext mining, when combined, can potentially influence both media houses and as well as their reader communities towards more fruitful and responsible political discourse online.

%Finally, of particular interest is analyzing the usage patterns of users belonging to different communities on a given website. It is likely communities correspond to groups of users commenting on the same thematic type of articles (for example, a cluster with a high comment rate on articles on gun violence, another on the economy).

\nocite{*}
\bibliographystyle{aaai}
\bibliography{News_Analysis.bib}

\begin{thebibliography}{}

\bibitem[\protect\citeauthoryear{Adamic and Glance}{2005}]{adamic2005political}
Adamic, L.~A., and Glance, N.
\newblock 2005.
\newblock The political blogosphere and the 2004 us election: divided they
  blog.
\newblock In {\em Proceedings of the 3rd international workshop on Link
  discovery},  36--43.
\newblock ACM.

\bibitem[\protect\citeauthoryear{Anderson \bgroup et al\mbox.\egroup
  }{2014}]{anderson2014nasty}
Anderson, A.~A.; Brossard, D.; Scheufele, D.~A.; Xenos, M.~A.; and Ladwig, P.
\newblock 2014.
\newblock The “nasty effect:” online incivility and risk perceptions of
  emerging technologies.
\newblock {\em Journal of Computer-Mediated Communication} 19(3):373--387.

\bibitem[\protect\citeauthoryear{Blei, Ng, and Jordan}{2003}]{blei2003latent}
Blei, D.~M.; Ng, A.~Y.; and Jordan, M.~I.
\newblock 2003.
\newblock Latent dirichlet allocation.
\newblock {\em Journal of machine Learning research} 3(Jan):993--1022.

\bibitem[\protect\citeauthoryear{Bradley}{1997}]{bradley1997use}
Bradley, A.~P.
\newblock 1997.
\newblock The use of the area under the roc curve in the evaluation of machine
  learning algorithms.
\newblock {\em Pattern recognition} 30(7):1145--1159.

\bibitem[\protect\citeauthoryear{Burnap and Williams}{2015}]{burnap2015cyber}
Burnap, P., and Williams, M.~L.
\newblock 2015.
\newblock Cyber hate speech on twitter: An application of machine
  classification and statistical modeling for policy and decision making.
\newblock {\em Policy \& Internet} 7(2):223--242.

\bibitem[\protect\citeauthoryear{Davidson \bgroup et al\mbox.\egroup
  }{2017}]{davidson2017automated}
Davidson, T.; Warmsley, D.; Macy, M.; and Weber, I.
\newblock 2017.
\newblock Automated hate speech detection and the problem of offensive
  language.
\newblock {\em arXiv preprint arXiv:1703.04009}.

\bibitem[\protect\citeauthoryear{DellaVigna and
  Kaplan}{2007}]{dellavigna2007fox}
DellaVigna, S., and Kaplan, E.
\newblock 2007.
\newblock The fox news effect: Media bias and voting.
\newblock {\em The Quarterly Journal of Economics} 122(3):1187--1234.

\bibitem[\protect\citeauthoryear{Gentzkow and
  Shapiro}{2006}]{gentzkow2006media}
Gentzkow, M., and Shapiro, J.~M.
\newblock 2006.
\newblock Media bias and reputation.
\newblock {\em Journal of political Economy} 114(2):280--316.

\bibitem[\protect\citeauthoryear{Groseclose and
  Milyo}{2005}]{groseclose2005measure}
Groseclose, T., and Milyo, J.
\newblock 2005.
\newblock A measure of media bias.
\newblock {\em The Quarterly Journal of Economics} 120(4):1191--1237.

\bibitem[\protect\citeauthoryear{Magu, Joshi, and
  Luo}{2017}]{magu2017detecting}
Magu, R.; Joshi, K.; and Luo, J.
\newblock 2017.
\newblock Detecting the hate code on social media.
\newblock In {\em Eleventh International AAAI Conference on Weblogs and Social
  Media}.

\bibitem[\protect\citeauthoryear{Nockleby}{2000}]{hateSpeechDef}
Nockleby, J.~T.
\newblock 2000.
\newblock Hate speech.
\newblock {\em Encyclopedia of the American constitution} 3:1277--79.

\bibitem[\protect\citeauthoryear{Papacharissi}{2004}]{papacharissi2004democracy}
Papacharissi, Z.
\newblock 2004.
\newblock Democracy online: Civility, politeness, and the democratic potential
  of online political discussion groups.
\newblock {\em New media \& society} 6(2):259--283.

\bibitem[\protect\citeauthoryear{Rieder}{2010}]{rieder2010no}
Rieder, R.
\newblock 2010.
\newblock No comment: It's time for news sites to stop allowing anonymous
  online comment.
\newblock {\em American Journalism Review} 32(2).

\bibitem[\protect\citeauthoryear{Shearer and Gottfried}{2017}]{shearer2017news}
Shearer, E., and Gottfried, J.
\newblock 2017.
\newblock News use across social media platforms 2017.
\newblock {\em Pew Res Cent}.

\bibitem[\protect\citeauthoryear{Wulczyn, Thain, and
  Dixon}{2017}]{wulczyn2017ex}
Wulczyn, E.; Thain, N.; and Dixon, L.
\newblock 2017.
\newblock Ex machina: Personal attacks seen at scale.
\newblock In {\em Proceedings of the 26th International Conference on World
  Wide Web},  1391--1399.
\newblock International World Wide Web Conferences Steering Committee.

\end{thebibliography}
\end{document}